\documentclass[rmp,twocolumn,byrevtex]{revtex4}
\usepackage{graphicx}
\usepackage{dcolumn}
\usepackage{bm}

\begin{document}
\title{A model for the temperature dependence of photoluminescence from self-assembled quantum dots\footnote{Submitted to Journal of Applied Physics: February 2006}}
\author{Bhavtosh Bansal\footnote{Electronic Address:
Bhavtosh.Bansal@fys.kuleuven.be}\footnote{Present Address:
INPAC-Institute for Nanoscale Physics and Chemistry, Pulsed Fields
Group, Katholieke Universiteit Leuven, Celestijnenlaan 200D,
Leuven B-3001, Belgium.}}
\affiliation{Department of Condensed Matter Physics and Materials
Science, Tata Institute of Fundamental Research, 1 Homi Bhabha
Road, Mumbai-400005, India.}
\date{\today}
\begin{abstract}
Photo-excited carriers, distributed among the localized states of self-assembled quantum dots, often show very anomalous
temperature dependent photoluminescence characteristics. The temperature dependence of the peak emission energy may be non-monotonic and
the emission linewidth can get narrower with increasing temperature. This paper describes a quasi-thermodynamic model that naturally explains these
observations. Specifically, we introduce a temperature dependent function to parameterize the degree of thermalization of carriers. This function
allows us to continuously interpolate between the well-defined low and high temperature limits of the carrier distribution function and describe the
observed anomalies in the photoluminescence spectra with just two fitting parameters. We show that the description is equivalent to assuming that the
partially thermalized carriers continue to be described by equilibrium statistics, but with a higher effective temperature.
Our treatment of the problem is computationally simpler than the usually employed rate equations based analyses [e.g. Sanguinetti, et al. Phys. Rev. B {\bf 60}, 8276 (1999)], which typically also have many more under-determined fitting parameters.
The model is extended to quantum dots with a bimodal size distribution.
\end{abstract}
\maketitle
\section{Introduction}
Photoluminescence(PL) spectroscopy has perhaps been the most
extensively used characterization tool for self-assembled quantum
dots (QD)\cite{nanooptoelectronics, sugawara}. This is due to the
relative ease of the measurement, the information it yields about
the extent of quantum confinement and inhomogeneous broadening of
the density of states and its very direct relevance in assessing
the use of these structures in QD lasers.

The temperature dependent PL spectra in self-assembled QD have
certain characteristic anomalies\cite{mackowski, polimeni, akiba,
hao lee, sanguinetti, Xu-epl, lobo}. At low temperatures, the
linewidth of emission generally decreases with increasing sample
temperature. The dependence of linewidth may also be non-monotonic
with a minimum at an intermediate temperature. Secondly, the
energy value corresponding to the peak of the emission spectrum
from an ensemble of dots is typically observed to decrease faster
with temperature than what is characteristic of either the bulk
material or individual dots. It is now well recognized that both these anomalies are due to the
non-thermal nature of the distribution of photoexcited
carriers\cite{polimeni, patane, runge}. Energetically distributed
bound states corresponding to QD of different sizes can act as
trapping centers for photoexcited carriers. These electron-hole
pairs,  excited high up in energy by the pump laser can rapidly
relax and get captured into the local potential minima in a random
way. At low temperatures, the finite recombination times may not
be long enough for these carriers to cross the local potential
minima and access the lowest energy (quasi-thermal) state, whereas
at higher temperatures the distribution can be expected to have
better thermalized.

Although these essential facts about the carrier dynamics
responsible for the temperature dependent anomalies, are well
understood\cite{runge}, modeling these processes, even in a
completely classical sense in terms of rate equation based
models\cite{sanguinetti}, has turned out to be rather cumbersome.
The rate equations\cite{bastard} attempt to extract out the most
important trapping and escape processes with a characteristic
activation energy associated with each\cite{sanguinetti, Xu-epl,
zhang}. While this approach is both physical and can yield
reasonable fits to the data, one typically requires a large number
of underdetermined parameters (rate constants, activation
energies, etc.) to explain the experimental data. Most of these
parameters may not be directly accessible to experimental
determination even in a time resolved measurement. In our opinion,
the study of carrier dynamics is best tackled by the more
microscopic quantum mechanical models\cite{runge}, while for
describing a steady state PL experiment something simpler should
suffice.

The purpose of this paper is to attempt a physically motivated
semi-empirical alternative that would allow for direct modelling
of the experimental data in terms of small number of fitting
parameters. We show that this is easily achieved by extending the
carrier thermalization model\cite{gurioli} for Stokes' shift in
mismatched alloy quantum wells. A qualitative
description\cite{polimeni, patane} of the picture already exists
in literature. Our model is equivalent to an equilibrium
description for localized carriers in terms of an effective
temperature.
\section{Theory}
\subsection{Formulation of the model}
Consider a semiconductor heterostructure sample, containing say,
$\sim 10^{10} cm^{-2}$ InAs quantum dots within GaAs matrix, that
is homogeneous on the macroscopic scale. The surface features
appear with a size distribution, which when unimodal, is generally
well described by a Gaussian function\cite{leonard, bhavtosh}.
Thus the mean lateral size $a_0$ and the variance $\sigma^2/2$
completely determine the morphology of a sample if, for
simplicity, all the dots are assumed to be of similar shape with
the same aspect ratio. Even within the assumption of a lack of
phase coherent coupling between different quantum dots, the
dependence of the ground state transition energy $E_t(a)$ on the
size and shape of the dot is rendered quite non-trivial for such
heterostructure samples. This is due to the finiteness of the
energy barriers at materials interface, non-spherical shapes and
the complicated strain distribution within the dot. Furthermore
since InAs is a `narrow gap' semiconductor and the confinement
energies can be even twice as much as the energy gap, one should
expect an active participation of many bands in determining the
energy levels. The problem of determining the size dependence of
the energy levels is therefore best treated
numerically\cite{grundmann}. Within a semi-empirical approach, one
may then attempt a fit to these numerically calculated transition
energies to  a single band effective mass-like equation, where the
confinement causes an increase in the effective energy gap, but
now with the exact values of the parameters determined by a
fit\cite{vijaysingh} to the more precise numerical results. For
instance, the size (basal length of the pyramid) dependence of the
transition energies as calculated by Grundmann et
al.\cite{grundmann} can be quite well described by a relation of
the kind
\begin{equation}\label{effective mass}
E_T(a)=E_g+A/a^k.
\end{equation} Fixing $E_g$ at 0.42 eV, the low temperature bandgap of InAs, $A=2.37$ eV nm$^{1/2}$ and $k=0.5$ provided a
good fit to the calculations in reference \cite{grundmann}.
Equipped with a relationship between the transition energies and
the dot sizes, one may immediately write down the expression for
the optical density of states and the absorption coefficient of
the ensemble as a sum of absorption by individual quantum dots.
The absorption coefficient $\alpha(E, a)$, corresponding to a
single dot is essentially the product of the optical matrix
element with the size dependent transition energy. The density of
states is therefore peaked at this resonance energy with a
Lorentzian \cite{huag koch} lineshape due to the finite lifetime
introduced by phonon and other scattering processes. This
homogeneous broadening $\gamma_e$ of the energies of individual
quantum dots is typically less than a millielectronvolt and is
completely drowned by the ensemble inhomogeneity effects which
give at least an order of magnitude larger contribution to the
optical density of states. One may conveniently take the limit
$\gamma_e \rightarrow 0$ and replace the Lorentzian by a delta
function. The dipole matrix element $|d_{cv}|$ would also be, in
general, dependent on the effective size of the exciton relative
to the quantum dot volume. We assume\cite{vijaysingh} that
$|d_{cv}|^2\sim a^{-m}$.  With these approximations, the ground
state interband absorption from the complete ensemble quantum dots with a Gaussian size distribution of mean $a_0$ and variance $\sigma^2/2$ may be
written as
\begin{equation}\label{definition:absorption coeff}
\alpha= C\int_0^\infty {a^{-m}} e^{-{(a-a_0)^2/\sigma^2}}
\delta[h\nu-E_T(a)]da.
\end{equation}
The constant $C$ clubs together the normalization of the
probability distribution function and the other size and energy
independent parameters.

The above integral is easily evaluated by
collapsing the delta function after making a suitable change of
variables and plugging in the relationship between the dot size
and ground state transition energy given by equation
\ref{effective mass}. For the time being one may proceed without
substituting the numerical values of $k$, $A$ and $E_g$ to keep
the treatment general. Integration of equation
\ref{definition:absorption coeff} yields (up to energy independent
constants)
\begin{equation}\label{eq:absorption coeff}
\alpha(E)\propto e^{-{1\over \sigma^2} \left[\left( {A\over
E-E_g}\right)^{1/k}-\left({A\over E_0-E_g}\right)^{1/k}\right]^2}
\left[{A\over E_0-E_g}\right]^\chi.
\end{equation}
$\chi=(k+1-m)/k$ and $E_0$ is the transition energy associated with the dot of size $a_0$.

In a typical low temperature non-resonant PL experiment, the
carriers are generated well above the transition energies,
typically within the region of the matrix and then they quickly
relax to the localized states within the quantum dots. In general,
one may expect that the trapping efficiency of larger dots may be
larger by the factor of $a^s$. This should be multiplied within
the integrand in equation \ref{definition:absorption coeff} with
the consequence that equation \ref{eq:absorption coeff} also
describes the {\em low temperature} PL spectra but with a new
value of $\chi=(k+1-m+s)/k$. This establishes the general form of
the low temperature PL spectra in terms of the quantum dot size
distribution and the corresponding exciton energies. At higher
temperatures, the PL spectra can significantly differ due to the
process of carrier thermalization within the localized states.
Under conditions of low excitation power and in the limit of high
temperatures, the emission spectra $I(E)$ is related to the
absorption spectrum by van Roosbroeck-Shockley type of
relation\cite{gurioli},
\begin{equation}
I(E)\propto \alpha(E, E_0) \exp(-E/k_BT),\;\;\;T\rightarrow\infty.
\end{equation}
$k_B$ is the Boltzmann constant and T the lattice temperature. The
proportionality constant in the above equation defines a
temperature dependent scale factor carrying information about
non-radiative pathways. It is evident that the emission spectra in
the low and high temperature limits differ essentially by only
this exponential factor, that which accounts for carrier
thermalization. Furthermore, the behavior between the two limits
is expected to be continuous and thus the essence of all carrier
dynamics is contained in the extent of carrier thermalization.
That is, one may just rewrite the exponent as
$exp[-\beta(T)E/k_BT]$ where the parameter $\beta(T)=0$ and $1$ at
low and high temperatures respectively. The temperature dependence
of the thermalization parameter, $\beta(T)$ contains the essence
of all the carrier dynamics through the multiple pathways. Within
the constraint of $0\leq \beta(T)\leq 1$ and $\beta(T_1)\leq
\beta(T_2)$, if $T_1\leq T_2$ the functional form of $\beta(T)$ is
not easy to estimate. For simplicity we may approximate it by a
sigmoidal function
\begin{equation}\label{eq. beta definition}
\beta(T)={1\over 1+ \exp[-(T-T_{th})/\Delta]}.
\end{equation}
$T_{th}$ and $\Delta$ are then the only two fit parameters of the
model. $T_{th}$ is the temperature when $\beta=0.5$ and $\Delta$
corresponds to the temperature interval over which the transition
from a `glassy' to a quasi-equilibrium state occurs. Then, the PL
spectrum at a temperature $T$ is
\begin{widetext}
\begin{eqnarray}{\label{pl final expression}}
I(E, T)\propto  \left[{A\over E-E_g}\right]^t 
e^{-\left[\left({A\over E-E_g}\right)^{1/k}-\left({A\over
E_0-E_g}\right)^{1/k}\right]^2/\sigma^2} e^{-{\beta E\over k_BT}},
\end{eqnarray}

with $t={[k+1-m+s(1-\beta)]/k}$. $s(1-\beta)$ is added to insure a
correct interpolation between the low and high temperature limits.
$s$ denotes exponent of the size dependent capture cross section
$\sim a^s$ of dots. $(1-\beta)$ ensures that the high temperature
PL is not explicitly dependent on the trapping efficiency of
individual dots. The maximum of this equation, corresponding to
the PL peak position at that temperature, is given by the solution
of the following equation:
\begin{eqnarray}
\left[{A\over E_p-E_g}\right]^{1/k}-\left[{A\over E_0-E_g}\right]^{1/k} 
={k\sigma^2\over 2}\left[{\beta\over k_BT}+ {t\over
E_p-E_g}\right]{(E_p-E_g)^{{1\over k}+1}\over A^{1/k}}
\end{eqnarray}
where $E_p$ is the temperature dependent peak energy.
\end{widetext}
While the values of $A$, $E_g$ and $k$ were discussed earlier, the
value of $t$ is yet to be determined. Within the effective mass
picture, the exciton Bohr radius in (bulk) InAs is of the order of
35 nm. This is a bit larger than the typical size of matured
InAs/GaAs dots (height 8 nm, pyramid shape, aspect ratio=5).  For
spherical dots, the oscillator strength $|d_{cv}|^2$ scales
inversely with volume\cite{kayanuma}, i.e. m=3, in the limit of
small dot sizes. Since our self-assembled dots are neither
spherical nor substantially larger or smaller than exciton size,
we do not have a clear understanding of the size dependence of the
matrix element. We therefore arbitrarily assume a value of $m$
that gives the simplest form to equation \ref{pl final
expression}, i.e. the value for which $k+1-m=0$, with the
understanding that the spectral shape of the absorption
coefficient, being largely determined by the exponential function,
is insensitive to this detail. Similarly, it is not easy to simply
estimate the volume dependence of the exciton capture cross
section by quantum dots and it is thus simplest to assume that the
process of trapping of excitons by quantum dots it is completely
random, independent of the quantum dot's volume. This is the
usually made assumption and in the present context leads to a
considerable simplification. With these
assumptions\cite{footnote1}, we have put $t=0$ in equation \ref{pl
final expression}. Therefore $E_0$ is the PL peak at zero
temperature, corresponding to the mean sized quantum dot in the
Gaussian ensemble. On the other hand, the shape of the absorption
spectrum itself is no longer Gaussian and gets skewed toward
higher energy, determined by the non-linear relationship between
the quantum dot size and transition energy.

There is also the temperature dependent contribution to the shift
in the peak of the PL spectrum due to the shrinkage of the gap
with temperature, which must be included in any analysis. Fig.
\ref{fig:other_possibilities} (dotted line) shows the temperature
dependence of the gap (with the energy zero arbitrarily shifted)
for InAs, with Varshni parameters ($a=0.000276 eVK^{-1}$ and
$\theta=93K$ in equation below) taken from literature and
therefore the position of the PL peak is given by
\begin{equation}\label{varshni}
E(T)=E_p(T)-{a T^2 \over \theta + T}.
\end{equation}

\subsection{Physical basis for the model}
Firstly we reiterate that the distribution function for
non-resonantly excited carriers within an ensemble of localized
states has two distinct regimes of behavior. A low temperatures,
the carrier distribution function is proportional to the total
density of available states. At very high temperatures, the
carriers are in thermal equilibrium with the lattice. In the
absence of cooperative effects, the crossover is expected to be
smooth. The temperature dependent function $\beta(T)$, which is
defined in Eq.\ref{eq. beta definition},  describes this
crossover.

The introduction of the parameter $\beta(T)$ amounts making the
effective carrier temperature different from the lattice
temperature while still assuming that the carriers are described
by an equilibrium distribution function but with a different
effective temperature. The value of the effective carrier
temperature is $T/\beta(T)$, always greater than or equal to the
lattice temperature. Thus the model is only approximate since it
aims at an effective equilibrium description for an essentially
non-equilibrium problem. In general, the idea of describing a
non-equilibrium distribution by a higher effective temperature is
not new. For example, this concept is extensively used in high
electric field electron transport studies in semiconductors. In a
related study of electron localization in
Ga$_{1-x}$In$_{x}$As$_{1-y}$N$_y$ quantum wells, we have recently
observed that the effective temperature of imperfectly thermalized
carriers is indeed higher the lattice temperature\cite{kadir}. The
concept of an effective electron temperature for localized
carriers in disordered quantum wells was previously proposed in
the paper by Gurioli, et al. \cite{gurioli} and further discussed
by Runge\cite{runge} (section 12a). Runge and
coworkers\cite{runge} have also used the equilibrium distribution
with an effective temperature to describe the results of their
simulations of the distribution of partially localized excitons.
Furthermore, their simulations also suggest an existence of two
distinct regimes, which appear in our model as temperatures above
and below $T_{th}$.

The temperature dependent behavior of the effective temperature
$T_{eff}$ may, at first, appear counter-intuitive. The effective
temperature is infinite both in the low and high temperature
limits. This is because at low temperatures, all the photoexcited
carriers are frozen into the states they first get captured into.
Thus the carrier distribution function essentially reflects the
total density of states. We recall that in the canonical ensemble,
temperature is defined in terms of probability of occupation P(E)
of a state of energy E through the relation,
\begin{equation}
P(E)={e^{-E/k_BT}\over Z}.
\end{equation}
Here $Z$ is the canonical partition function. So it follows that
every state has an equal probability of occupation $P(E)=1/Z$ when
the temperature is infinite. This is consistent with the condition
we had imposed earlier (in an apparently arbitrary manner) that
$\beta=0$ for carriers to be uniformly distributed over the
complete density of states.

It is also tempting to compare our situation with that encountered
in the glass transition\cite{glassy_reference}. In context of
glasses, the high effective temperature corresponds to an extra
configurational entropy contribution. The notion of a crossover
temperature is also well-established for glasses and is defined as
the temperature at which certain degrees of freedom get frozen. A
more microscopic guess for the function form of the parameter
$\beta$ would require modeling of the barriers associated with the
basins of attraction for carriers. Very roughly, the parameters
$T_{eff}$ and $\Delta$ can be thought to correspond to the glass
transition temperature and {\it fragility} respectively.

Finally, we emphasize that the treatment presented in this paper
is restricted to the description of the temperature dependent
emission from the ground states of quantum dots.
\section{Results and Discussion}
\subsection{Dots with unimodal size distribution}
\begin{figure}[!h]
\begin{center}
\resizebox{!}{10cm} {\includegraphics{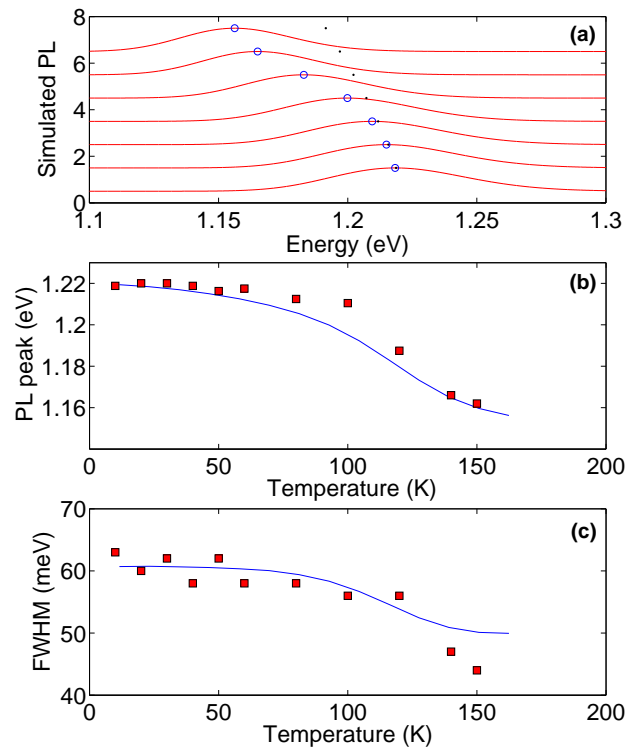}}
\caption{\label{fig:sanguinetti_comp}{\it (a) Simulated
photoluminescence spectra at different temperatures,
11K(bottom)-160K (top) in equally spaced temperature intervals.
(dots) the expected peak position assuming only the temperature
dependent shift in the gap and (circles) the actual peak position.
The energy difference between the two corresponds to the
Stokes'-like shift (b)Simulated (solid line) and experimental
(circles) PL peak energies as a function of temperature.
(c)Simulated (solid line) and experimental PL full-widths at half
maximum as a function of temperature. The experimental data was
taken from reference \cite{sanguinetti} where the same data is
fitted to an alternate model. The theoretical fit depicted here is
with the values of $\sigma=0.8 nm$, $E_0=1.22 eV$,
$T_{th}=127K$,\,\,$\Delta=15K$.}}
\end{center}
\end{figure}

With the above mentioned assumptions, the normalized PL spectrum
is described by the following simple equation ($t=0$ and $k=0.5$
in equation \ref{pl final expression})
\begin{equation}\label{unimodal_PL}
I(E)=I_0(T) e^{-{1\over \sigma^2} \left[\left( {A\over
E-E_g}\right)^2-\left({A\over E_0-E_g}\right)^2\right]^2}
e^{-{\beta E \over k_BT}}
\end{equation}
with $\beta(T)$, defined in equation \ref{eq. beta definition}, as
the fitting parameter.

To compare with experiments, the energy dependence of PL intensity
in equation \ref{unimodal_PL} must be of course also be translated
along the energy axis by an amount suggested by equation
\ref{varshni}. $I_0(T)$ denotes the temperature dependent scale
factor. While this has been previously studied and
modelled\cite{Le Ru}, we do not discuss it further because one
usually observes the rather predictable Arrhenius-type activation
which may as well be put in by hand. Furthermore, the
non-radiative decay channels may depend on a particular sample's
past history (e.g., the growth route employed to prepare the
sample and the nature of defects) and thus not general enough to
be included in a model like this.

Fig.\ref{fig:sanguinetti_comp} shows a comparison of our
calculations with the published results of Sanguinetti et al.
(figure 4 in reference \cite{sanguinetti}). The assumption of
$\beta\approx 0$ at the lowest temperature fixed $\sigma=0.8nm$
and $E_0=1.219eV$. Then the rest of the fit was accomplished with
the values of $\Delta=15K$ and $T_{th}=127K$. This fitting
procedure is highly constrained since it requires two separate
curves  to simultaneously fit on the basis of just two free
parameters. This should be compared with the rate equation based
models, reference \cite{sanguinetti} for example had eight fitting
parameters.

\begin{figure}[!h]
\begin{center}
\resizebox{!}{10cm} {\includegraphics{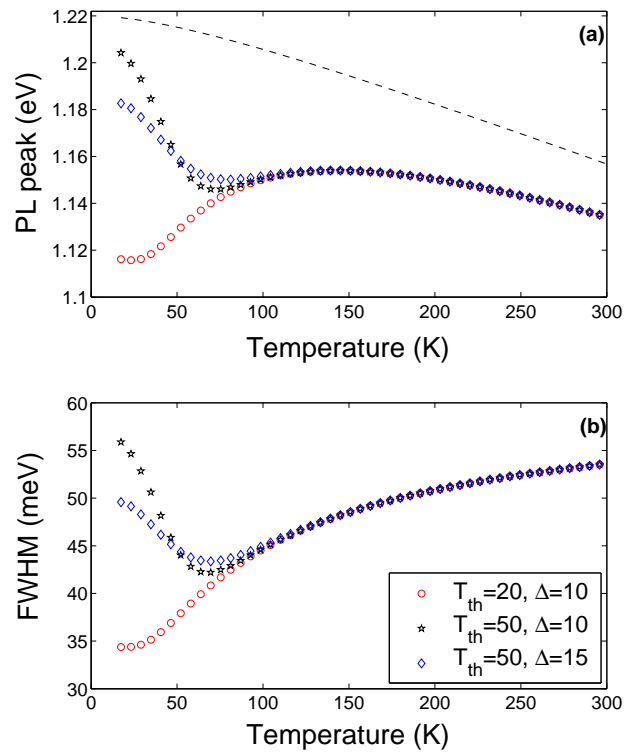}}
\caption{\label{fig:other_possibilities}{\it Calculated
temperature dependent characteristics of the PL spectra (a)peak
positions (b)FWHM for different values of $T_{th}$ and $\Delta$
(in units of Kelvin) with for a fixed values of $\sigma=0.8nm$ and $E_0=1.22 eV$. The
dotted line in subfigure (a) depicts the temperature dependence of
bulk InAs with the energy zero arbitrarily shifted.}}
\end{center}
\end{figure}
The temperature dependent behaviour of FWHM and peak positions for
other values of $T_{th}$ and $\Delta$ is depicted in
Fig.\ref{fig:other_possibilities}. For ease of comparison, the
ensemble characteristics, $E_0$ and $\sigma$ are the same as those
used to generate Fig.\ref{fig:sanguinetti_comp}, and only $T_{th}$
and $\Delta$ are varied. The model yields a variety of trends at
low temperatures. Some of these tends have not only been
experimentally observed in the PL from self-assembled quantum dots
but seem to be a general characteristic of emission from any
inhomogeneous ensemble of localized states. Thus it is tempting to
also use the present model to analyse the PL from localized states
in the disordered quantum wells as well, in particular the
recently much studied dilute nitrides (for a review see e.g.,
reference \cite{dilute nitrides}) and InGaN (see e.g.,
\cite{yu_apl}) quantum wells.

But to extend this model to localized states in quantum wells
would require us to guess an appropriate form of the density of
states, which would be a combination of localized band-tail and
higher energy extended states. In contrast, for self-assembled
quantum dots, all states can be considered to be localized and the
density of states can be more or less reliably extracted from the
size distribution which in turn is directly accessible to surface
probe microscope analysis. Furthermore, the origin of low
temperature PL in disordered quantum wells is different from that
in quantum dots. While in quantum dots, the PL spectrum at zero
temperature can be taken to be proportional to the {\em total}
density of states (all of which are localized and delta-function
like), only a small fraction (corresponding to the local potential
energy minima in a classical sense\cite{yang} or the band tail
states below the mobility edge) of the total density of states
contribute to the low temperature PL in disordered quantum wells.

It is also important to mention that a significant advantage of an
effective equilibrium description is that one can continue to use
the results from equilibrium theory\cite{eliseev_jap, eliseev_apl}
by just replacing the temperature by effective temperature
$T_{eff}$. The non-monotonic behavior of the linewidths just
follows the non-monotonicity of the effective temperature, since
the photoluminescence linewidths in equilibrium theory are
proportional to $k_BT$. Similarly the temperature dependent
``Stokes shift"\cite{eliseev_jap} is described simply by
$\sigma^2/k_BT_{eff}\sim (k_BT_{eff})^2/k_BT_{eff}\sim
k_BT_{eff}$. This is evident in Fig.
\ref{fig:other_possibilities} where the curves in subfigures (a)
and (b) seem to follow each other. Also note that our results are
identical to the results in reference \cite{eliseev_apl} in the
high temperature limit of $\beta=1$.
\subsection{Dots with bimodal size distribution}
QD ensembles often have a bimodal size distribution, especially
during the intermediate stage of growth. This size bimodality also
reflects in the low temperature PL spectra with the appearance of
two peaks corresponding to the two QD families. While there have
been many experimental studies of the temperature dependence of
the two peaks\cite{zhang, saint-girons, Zhang-huang-jcg, cm-lee,
wang-jcg, kissel}, the quantitative theoretical understanding so
far has been limited to only the change in the relative intensity
of the two peaks as a function of temperature\cite{zhang,
saint-girons}. This was done again on the basis of rate equations
corresponding to the change in the carrier population in the
quantum dots belonging to the two families.
\begin{figure}[!h]
\begin{center}
\resizebox{!}{11cm} {\includegraphics{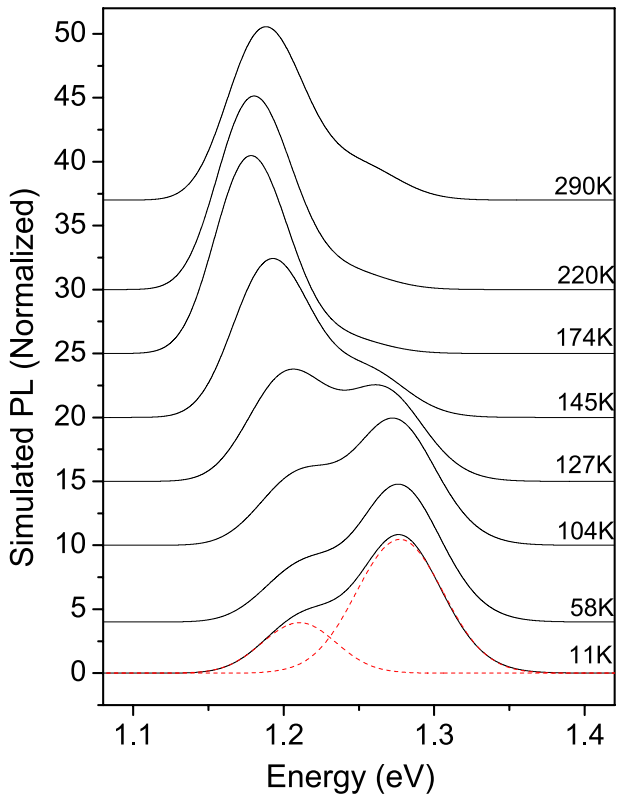}}
\caption{\label{fig:bimodal_pl_simulated}{\it (Solid line)
Simulated photoluminescence spectra at different temperatures for
a bimodally size distributed quantum dots ensemble. The curves are
normalized such that the total area of each is the same. (Dashed
line) 11K spectrum is fitted to two Gaussians. All other curves
are also well described as a sum of two Gaussians. Note that the
higher energy peak that virtually disappears at 220K is again
visible in the 290K spectrum.}}
\end{center}
\end{figure}
Our model is easily extended to describe PL in  bimodally size
distributed dots, by simply accounting for the modified size
distribution which is now a sum of two Gaussian functions, each
with a specific mean size and size dispersion around that mean.
Denoting by $E_{0_{1,2}}$, the energies corresponding to the two
mean sizes and the ($\sigma_1$ and $\sigma_2$) the size
dispersions in the two families, we may redefine the optical
density of states which we had assumed to be proportional to the
absorption coefficient which in turn was proportional to the PL
spectra measured at the lowest temperature as
\begin{eqnarray}
\alpha_{\rm{bimod}} (E)=  \sum_{i=1,2} \alpha_{0_{i}} e^{-{1\over
\sigma_i^2} \left[\left( {A\over E-E_g}\right)^2-\left({A\over
E_{0i}-E_g}\right)^2\right]^2}.
\end{eqnarray}
Here the subscripts $1$ and $2$ label the two families of dots and
the ratio of the constants $\alpha_{0_{1,2}}$ is just the relative
density of dots in the two families. With this new definition of
$\alpha_{\rm{bimod}}$, and keeping the rest of the treatment the
same, we have plotted simulated PL spectra at different
temperatures in Fig.\ref{fig:bimodal_pl_simulated}. The following
set of parameter values was used: $\sigma_1=0.8$nm,
$\sigma_2=0.58$nm, $E_{01}=1.22$eV, $E_{02}=1.28$eV,
$T_{th}=150K$, $\Delta=15K$. The ratio of the peak intensities was
fixed to $I(E_{01}):I(E_{01})= 2:1$ at zero temperature. All these
values were more or less arbitrarily chosen and supposed to be
representative of the `typical' temperature dependent behavior of
PL from bimodally size distributed dots.

\begin{figure}[!h]
\begin{center}
\resizebox{!}{11cm}
{\includegraphics{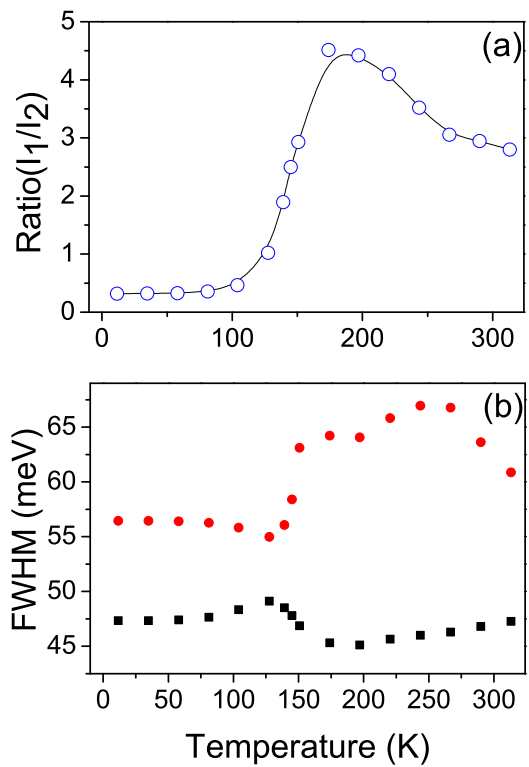}}
\caption{\label{fig:intensity_ratio_fwhm}{\it Analysis of
simulated PL spectra, Fig.\ref{fig:bimodal_pl_simulated}. (a)
Temperature dependence of the ratio of the integrated intensity of
the two peaks. The spectrum shows a carrier transfer to lower
energy peak corresponding to larger dots at intermediate
temperatures. In accordance with
Fig.\ref{fig:bimodal_pl_simulated}, the peak reappears at higher
temperatures. Note its qualitative similarity with Fig.4 in the
work of Zhang, et al.\cite{zhang}. (b) FWMH of the two peaks at
different temperatures. }}
\end{center}
\end{figure}

The spectra in Fig.\ref{fig:bimodal_pl_simulated} may be compared
with the experimentally measured temperature dependent line shapes
from literature, for example Fig.1 from reference
\cite{saint-girons}, Fig.2 in reference \cite{cm-lee} and Fig.3 of
reference\cite{zhang}.

We note that the experimentally observed trends are reproduced.
For ease of analysis, we again ignore the effects of carrier loss
to non-radiative decay channels and focus only on the effects of
carrier redistribution. Therefore, the curves in
Fig.\ref{fig:bimodal_pl_simulated} are plotted such that the total
integrated intensity in same in each case. The spectra in
Fig.\ref{fig:bimodal_pl_simulated} between 11K-150K show the
transfer of carriers from smaller to larger dots. The intensity of
the higher energy peak is observed to decrease. Since $T_{th}$ is
the simulations was 150K, we observe that the thermalization of
carriers is nearly complete by 200K. This leads to all the
carriers being preferentially in larger dots and a virtual
disappearance of the high energy peak. The carrier transfer from
the smaller to larger quantum dots was found to be activated.
Interestingly, at high enough temperatures, we observe that the
high energy peak re-appears when the thermally broadened carrier
distribution is so smeared that carriers have sufficient thermal
energy to reside on smaller dots. This (somewhat unexpected)
behavior has previously been observed in two contexts. Firstly,
the plots of the ratio of the intensities between to the two peaks
(Fig.\ref{fig:intensity_ratio_fwhm} (a)) is qualitatively very
similar to the Fig.4 of reference\cite{zhang}. Note that the
analysis by Zhang, et al.\cite{zhang} was based on rate equations
and employed many more parameters to simulate a curve similar to
Fig.\ref{fig:intensity_ratio_fwhm}(a) and had no predictions about
the linewidths. Secondly, the reappearance of the high energy
feature at high temperature has also been observed in context of
pronounced appearance of the wetting layer peak\cite{patane-wl} in
the high temperature PL spectra from InAs quantum dots.

Finally, the temperature dependence of the positions and the
linewidths of the two Gaussians are plotted in
Fig.\ref{fig:intensity_ratio_fwhm}(b) which shows a small but
systematic decrease in the linewidth for the higher energy peak
and a corresponding increase in the lower energy peak's linewidth.
This feature is generic to the bimodal dots' PL. After 150K, we
observe an interesting anomaly as the behavior of the linewidths
reverses. This `anomaly' has also been experimentally
observed\cite{cm-lee} although not satisfactorily explained.

\section{Summary}
We have presented an extremely simple model for the observed
anomalies in the temperature dependence of the PL spectra of
self-assembled quantum dots. Unlike the previously proposed
methods of analysis, we do not invoke a rate equations to model
the steady state carrier dynamics but instead introduce a
physically motivated factor that parameterizes the degree of
thermalization of photoexcited carriers. Since the low and high
temperature limits of the carrier distribution are well
understood, the process of interpolation allows for an easy
visualization of the carrier distribution at different
temperatures in terms of this single parameter. Using this model,
we could simulate a variety of experimentally observed trends for
quantum dots ensembles with both a unimodal and a bimodal
distribution.
\section{Acknowledgements}
I thank B.M. Arora for introducing me to self-assembled quantum
dots and PL measurements and for many discussions on these topics.
I also thank Jayeeta Bhattacharya, Sandip Ghosh and K. L.
Narasimhan for very useful comments.


\begin{references}
\bibitem{nanooptoelectronics}M. Grundmann (Ed.) {\it Nano-Optoelectronics}, (Springer-Verlag, Berlin 2002).
\bibitem{sugawara}M. Sugawara (Ed.) {\it Semiconductors and Semimetals}, vol 60, Academic 
Press, San Diego (1999).
\bibitem{mackowski}S. Mackowski, G. Prechtl, W. Heiss, F. V. Kyrychenko, G. Karczewski, and J. Kossut, Phys. Rev. B {\bf 69}, 205325 (2004).
\bibitem{akiba}Keiichirou Akiba, Naoki Yamamoto, Vincenzo Grillo, Akira Genseki, and Yoshio Watanabe, Phys. Rev. B {\bf 70}, 165322 (2004).
\bibitem{polimeni}A. Polimeni, A. Patane, M. Henini, L. Eaves, and P. C. Main,
Phys. Rev. B {\bf 59}, 5064 (1999).
\bibitem{lobo}C. Lobo, N. Perret, D. Morris, J. Zou, D. J. H. Cockayne, M. B. Johnston, M. Gal, and R. Leon, Phys. Rev. B {\bf 62}, 2737 (2000).
\bibitem{hao lee}Hao Lee, Weidong Yang, and Peter C. Sercel, Phys. Rev. B {\bf 55}, 9757 (1997).
\bibitem{sanguinetti}S. Sanguinetti, M. Henini, M. Grassi Alessi, M. Capizzi, P. Frigeri, and S. Franchi, Phys. Rev. B {\bf 60}, 8276 (1999)
\bibitem{Xu-epl}Q. Li, S. J. Xu, M. H. Xie, and S. Y. Tong, Europhys. Lett. {\bf 71},
994 (2005).
\bibitem{zhang} Y. C. Zhang, C. J. Huang, F. Q. Liu, B. Xu, J. Wu, Y. H. Chen, D. Ding, W. H. Jiang, X. L. Ye, and Z. G. Wang,  J. Appl. Phys. {\bf 90}, 1973 (2001).
\bibitem{patane}A. Patane, A. Levin, A. Polimeni, L. Eaves, P. C. Main, and M. Henini, G. Hill, Phys. Rev. B {\bf 62}, 11084 (2000).
\bibitem{runge}E. Runge, in {\it Solid State Physics}, edited by H. Ehrenreich and F. Spaepen, (Academic Press, San Diego, 2002), Vol. 57, p. 150.
\bibitem{bastard}For a general introduction to rate equations, see for example, G. Bastard, {\em Wave Mechanics Applied to Semiconductor Heterostructures},  (Les Editions de Physique, Les Ulis, France, 1992).
\bibitem{gurioli}M. Gurioli, A. Vinattieri, J. Martinez-Pastor, and M. Colocci, Phys. Rev. B {\bf 50}, 11817 (1994).
\bibitem{leonard}D. Leonard, K. Pond, and P. M. Petroff, Phys. Rev. B {\bf 50}, 11687 (1994).
\bibitem{bhavtosh}B. Bansal, M. R. Gokhale, A. Bhattachraya and B. M. Arora, Appl. Phys. Lett. {\bf 87}, 203104 (2005).
\bibitem{grundmann}M. Grundmann, O. Stier, and D. Bimberg, Phys. Rev. B {\bf 52}, 11969 (1995).
\bibitem{vijaysingh}V. Ranjan, V. A. Singh and G. C. John, Phys. Rev. B {\bf 58}, 1158
(1998).
\bibitem{huag koch}H. Huag and S. W. Koch, {\it Quantum Theory of the Optical and Electronic Properties of Semiconductor} (World Scientific, Singapore, 1998).
\bibitem{kayanuma}Y. Kayanuma, Phys. Rev. B {\bf 38}, 9797 (1988).
\bibitem{footnote1}Although, we have, to keep things simple and in absence of definitive information assumed $t=0$,
the parameter $t$ is in principle measurable through 'Stokes
shift' between the peaks of the absorption and the PL spectra.
\bibitem{kadir}B. Bansal, A. Kadir, A. Bhattachraya, B.M. Arora, and R. Bhat, Appl. Phys. Lett. {\bf 89}, 032110 (2006); cond-mat/0511282.
\bibitem{glassy_reference}For example, Th. M. Nieuwenhuizen, Phys. Rev. Lett. {\bf 80}, 5580
(1998). P. G. Debenedetti and F. H. Stillinger, Nature {\bf 410},
259 (2001).
\bibitem{Le Ru}E. C. Le Ru, J. Fack, and R. Murray, Phys. Rev. B {\bf 67}, 245318 (2003).
\bibitem{dilute nitrides}P. J. Klar, Prog. Sol. Stat. Chem., {\bf 31}, 301 (2003).
\bibitem{yu_apl}K. L. Teo, J. S. Colton, P. Y. Yu, E. R. Weber, M. F. Li, W. Liu, K. Uchida, H. Tokunaga, N. Akutsu, and K. Matsumoto
Appl. Phys. Lett. {\bf 73}, 1697 (1998).
\bibitem{yang}F. Yang, M. Wilkinson, E. J. Austin, and K. P. O'Donnell, Phys. Rev. Lett. {\bf 70}, 323 (1993), ibid. {\bf 72}, 1945 (1994) (Erratum).
\bibitem{eliseev_jap}P. G. Eliseev, J. Appl. Phys. {\bf 93}, 5404 (2003).
\bibitem{eliseev_apl}P. G. Eliseev, P. Perlin, J. Lee, M. Osinski, Appl. Phys. Lett., {\bf 71}, 569 (1997).
\bibitem{Zhang-huang-jcg}Y. C. Zhang, C.J. Huang, F.Q. Liu, B. Xu, D. Ding, W.H. Jiang, Y.F. Li,
X.L. Ye, J. Wu, Y.H. Chen, and Z.G. Wang, J. Crystal Growth {\bf
219}, 199 (2000).
\bibitem{saint-girons}G. Saint-Girons and I. Sagnes, J. Appl. Phys. {\bf 91}, 10115 (2002).
\bibitem{cm-lee}C.M. Lee, S.H. Choi, J.C. Seo,  J.I. Lee,  J.Y. Leem, and I.K. Han, J. Korean Phys. Soc. {\bf 45} 1615 (2004).
\bibitem{wang-jcg}H.L. Wang, D. Ning, and S.L. Feng, J. Crystal Growth {\bf 209}, 630 (2000).
\bibitem{kissel} H. Kissel, U. Muller, C. Walther, W.T. Masselink, Yu I. Mazur, G.G. Tarasov, and M.P. Lisita, Phys. Rev. B, {\bf 62}, 7213 (2000).
\bibitem{patane-wl}A. Patan\'e, A. Polimeni, P. C. Main, M. Henini, and L. Eaves, Appl. Phys. Lett. {\bf 75}, 814 (1999).
\end{references}
\end{document}